\documentclass{emulateapj}

\usepackage{amsmath,epsf,natbib,graphicx,color}
\usepackage{ulem}
\newcommand{\sigT}{\mbox{$\sigma_{\mbox{\tiny T}}$}}
\newcommand{\Tcmb}{\mbox{$T_{\mbox{\tiny CMB}}$}}
\newcommand{\kB}{\mbox{$k_{\mbox{\tiny B}}$}}

\newcommand{\rhogas}{\mbox{$\rho_{\mbox{\scriptsize gas}}$}}
\newcommand{\rhotot}{\mbox{$\rho_{\mbox{\scriptsize tot}}$}}
\newcommand{\Mgas}{\mbox{$M_{\mbox{\scriptsize gas}}$}}
\newcommand{\Mtot}{\mbox{$M_{\mbox{\scriptsize tot}}$}}
\newcommand{\Yint}{\mbox{$Y_{\mbox{\scriptsize int}}$}}

\newcommand{\Ysph}{\mbox{$Y_{\mbox{\scriptsize sph}}$}}
\newcommand{\fgas}{\mbox{$f_{\mbox{\scriptsize gas}}$}}
\newcommand{\LCDM}{\mbox{$\Lambda$CDM}}
\newcommand{\Pe}{\mbox{$P_{\mbox{\scriptsize e}}$}}

\begin{document}
\submitted{Published 2011 January 31; this version includes a surface pressure correction}
\title{A new approach to obtaining cluster mass from Sunyaev--Zel'dovich Effect
observations}

\author{Tony~Mroczkowski,\altaffilmark{1,2}}
\altaffiltext{1}{Department of Physics and Astronomy, 
University of Pennsylvania, Philadelphia, PA 19104, USA}
\altaffiltext{2}{Einstein Postdoctoral Fellow}

\begin{abstract} 
The accurate determination of cluster total mass is crucial for their use as 
probes of cosmology.  Recently, the Sunyaev--Zel'dovich effect (SZE) has been 
exploited in surveys to find galaxy clusters, but X-ray or lensing follow-up observations,
or empirically determined scaling relations between SZE flux and total mass, 
have been required to estimate their masses.
Here, we demonstrate a new method of mass determination from SZE observations, 
applicable in the absence of X-ray or lensing data.  This method relies on the virial 
relation and a minimal set of assumptions, following an approach 
analogous to that used for stellar structure.
By exploiting the virial relation, we implicitly incorporate an additional 
constraint from thermodynamics that is not used in deriving the equation 
of hydrostatic equilibrium.  This allows us to relate cluster total mass directly
to the robustly-determined quantity, the integrated SZE flux.
\end{abstract}


\section{Introduction}

Clusters of galaxies are thought to be the largest gravitationally-bound 
objects in the universe and therefore good tracers of cosmology.
Ongoing cluster surveys, such as those with the Atacama Cosmology Telescope
\citep[ACT;][]{kosowsky2003,menanteau2010}, the South Pole Telescope
\citep[SPT;][]{ruhl2004,plagge2010}, and Planck \citep{rosset2010,Planck2011ESZ,Planck2011scaling}, 
have the potential to place tight constraints on cosmological parameters 
with the clusters they discover.  
These surveys utilize the Sunyaev--Zel'dovich effect (SZE), which
has redshift independent surface brightness and
arises by Compton scattering of cosmic microwave background (CMB) photons off 
of the hot electrons in clusters of galaxies \citep{zeldovich1969,sunyaev1972}.  
However, the interpretation of cluster yields relies on the accurate
determination of the scaling between integrated SZE flux and cluster
total mass \citep{mccarthy2003a,motl2005,nagai2006}. 

Because the SZE intensity varies as the line-of-sight integral of
thermal electron pressure, a cluster's total, integrated SZE flux
scales linearly with the volumetric integral of thermal pressure,
which is thermal energy (see, e.g., Equation \ref{eq:Ethermal}).  
To the extent that clusters are virialized and 
supported hydrostatically by thermal pressure, a cluster's total SZE flux
will closely track its gravitational energy, thereby motivating SZE
flux as a proxy for cluster mass.
Approaches exploiting this expected tight correlation have traditionally 
relied on empirical relations between SZE flux and total mass determined 
from X-ray observations or optical lensing studies.

We present a new approach for mass determination from SZE data alone, applicable
in the absence of X-ray or lensing data, that exploits the virial relation 
and a minimal amount of simplifying assumptions about cluster astrophysics.
In Section \ref{szemass} we describe how SZE observations can constrain 
thermal energy, and we relate this directly to cluster total mass via the virial theorem.  
In Section \ref{szeobs}, we demonstrate this mass determination on previously
published SZE data.  In Section \ref{conc}, we assess how these simplifying assumptions 
impact this method and offer conclusions.

\section{Thermal Energy Constraints from Observations of the Sunyaev--Zel'dovich Effect}\label{szemass}

The thermal SZE is a small ($\lesssim 10^{-3}$) distortion 
in CMB intensity caused by inverse Compton scattering of CMB 
photons by energetic electrons in the hot intracluster medium
\citep[ICM;][]{zeldovich1969,sunyaev1972}.  This spectral distortion can be
expressed, for dimensionless frequency $x \equiv h\nu/\kB \Tcmb$, where
 $h$ is Planck's constant, $\nu$ is frequency, \kB\ is Boltzmann's constant, and 
\Tcmb\ is the primary CMB temperature, as
the change in intensity $\Delta I_{\rm SZE}$ relative to the primary CMB intensity 
normalization $I_0$,
\begin{eqnarray}
\label{eq:thermal_sz}
\frac{\Delta I_{\rm SZE}}{I_0} &=& g(x,T_e) ~ y.
\end{eqnarray}
The factor $g(x,T_e)$ in Equation \ref{eq:thermal_sz} encapsulates the frequency dependence 
of the SZE intensity. For non-relativistic electrons, 
\begin{equation}
\label{eq:g_x}
g(x) =  \frac{x^4 e^x}{(e^x-1)^2} \left(x \frac{e^x + 1}{e^x - 1} - 4\right).
\end{equation}
At low frequencies ($\lesssim 600~\rm GHz$), relativistic corrections to Equation \ref{eq:g_x} are fairly 
straightforward to apply \citep[see, e.g.,][]{itoh1998}.
The Compton $y$ parameter in Equation \ref{eq:thermal_sz} is defined as
\begin{equation}
\label{eq:compy}
y \equiv \frac{\kB \, \sigT}{m_{\rm e} c^2} \int \! n_e T_e \,d\ell = 
\frac{\sigT}{m_{\rm e} c^2} \int \! P_e \,d\ell,
\end{equation}
where \sigT\ is the Thomson scattering cross-section of the electron,
$\ell$ is the line of sight, and $m_{\rm e} c^2$ is an electron's rest energy, and
the primary CMB intensity normalization is
$I_0 = 2 (\kB \Tcmb)^3 (h c)^{-2} = 2.7033 \times 10^8~\rm Jy~Sr^{-1}$.
Note that we have assumed the ideal gas law ($P_e = n_e \kB T_e$) in Equation \ref{eq:compy}
to relate electron pressure $P_e$ to electron number density $n_e$ and temperature $T_e$.

From Equation \ref{eq:compy}, one can see that resolved observations of the thermal SZE 
from a cluster can be used to constrain its electron pressure profile $\Pe(r)$.  
This can be related to the total pressure 
as $P_{\rm gas}=(1+1/\mu_{\rm e}) \Pe$, where $\mu_{\rm e} = 2/(1+X)$ is the mean particle 
weight per electron and $X$ is the mass fraction of hydrogen. 
While deep X-ray observations have shown that the distribution of heavy elements 
in the ICM varies with radius \citep[e.g.][]{vikhlinin2005a,peterson2001}, 
and theoretical studies indicate that helium sedimentation into the cluster core 
will also impact $\mu_{\rm e}$ \citep[e.g.][]{markevitch2007,peng2009}, 
we make the simplifying assumption that $\mu_{\rm e} = 1.17$. 
We assess the impact of this assumption in Section \ref{conc}.

A common SZE observable used in SZE mass scaling relations is \Yint, 
the Compton $y$ parameter integrated over some region of the sky, defined as
\begin{equation}
\label{eq:Yint}
\Yint \equiv \int \! y \, d\Omega.
\end{equation}
Because \Yint\ is proportional to the surface brightness of the SZE integrated over
a region of the sky, it tracks the integrated cluster SZE flux.
For a spherically symmetric electron pressure profile $\Pe(r)$, the spherically-integrated 
version of \Yint\ is \citep[e.g.,][]{mroczkowski2009}
\begin{eqnarray}
\label{eq:Ysph}
\Ysph(r) 
&\equiv& \frac{\sigT}{m_{\rm e} c^2} \int_{0}^{r} \!\! \Pe (r') \, 4 \pi  r'^2 dr' \\
\label{eq:Ysph2}
&=& \frac{\sigT}{(1+1/\mu_{\rm e})\, m_{\rm e} c^2} \int_{0}^{r} \!\! P_{\rm gas} (r') \, 4 \pi r'^2 dr' \\
\label{eq:Ysph3}
&=& \frac{2 \sigT E_{\rm th}(r)}{3 (1+1/\mu_{\rm e})\, m_{\rm e} c^2}.
\end{eqnarray}
We have used the fact in Equation \ref{eq:Ysph3} that the thermal energy within $r$ is
\begin{equation}
\label{eq:Ethermal}
E_{\rm th}(r) = \frac{3}{2}\int_0^{\, r} \! P_{\rm gas}(r^\prime) \, 4 \pi  r^{\prime 2} dr^\prime
\end{equation}
for a monatomic, ideal gas.

Through Equation \ref{eq:Ysph3}, one can see that the SZE observable \Ysph\ relates
directly to the thermal energy content of the cluster.
To the extent that a cluster is virialized and supported by thermal pressure, 
this quantity will closely track the gravitational energy $U_g(r)$ via the virial relation,
\begin{equation}
\label{eq:vir}
2 E_{\rm th}(r) -3P(r) V = -U_g(r).
\end{equation}
{The $-3PV$ term, where $V=4\pi r^3/3$ is the volume at $r$, accounts for the non-vanishing
surface pressure and -- as noted recently by Colin Hill -- must be taken into account when solving
for the mass.
This term works to reduce the amount of gravitating mass required to hold the gas within $r$.}
Using Equation \ref{eq:Ethermal}, which derives from statistical mechanics, 
to relate pressure to thermal energy, we note that the virial relation (Equation \ref{eq:vir}) is derived 
(see, e.g., \cite{schwarzschild1958} or \citet{KippenhahnWeigert1990}, who derive and discuss
the surface pressure term) from the equation of hydrostatic equilibrium (HSE), 
\begin{eqnarray}
\label{eq:hse}
\frac{dP_{\rm gas}}{dr} &=& - \rhogas(r) \frac{G \Mtot}{r^2},
\end{eqnarray}
where $\rhogas(r)$ is the gas density as a function of radius $r$, 
$\Mtot(r)$ is the total mass within $r$, and $G$ is the gravitational constant.
The equation of HSE is derived from fluid mechanics, specifically the 
equations of motion and continuity \cite[see, e.g.,][]{schwarzschild1958, sarazin1988}.
Mass estimates based on HSE traditionally assume a spherically-symmetric 
mass distribution and the ideal gas law, and therefore adopting the virial 
relation is no more restrictive than the assumptions typically required for HSE 
mass determinations.  
However, Equation \ref{eq:Ethermal} provides an additional, key constraint not used in pure HSE 
mass determinations, which (assuming the ideal gas law) only require two of these 
three ICM profiles: density, temperature, and pressure.
As we discuss in \S \ref{conc}, the virial mass estimate is proportional to
the square root of scalar changes in the pressure profile,
while the same changes would result in linear changes in the HSE mass estimate, as
can be seen by examining Equation \ref{eq:hse}.

Adopting the Navarro, Frenk, and White profile \citep[][hereafter NFW]{navarro1997} 
to describe the total mass distribution (i.e., baryonic + dark matter distribution), 
the total density is radially distributed as
\begin{equation}
\label{eq:nfw}
\rhotot(r) =  \frac{\rho_0}{(r/R_s) (1+r/R_s)^{2}},
\end{equation}
and the total mass within $r$ is
\begin{equation}
\label{eq:Mtot}
\begin{split}
\Mtot(r) &= \int_0^{\, r}  \!\! \rhotot(r^\prime) \, 4 \pi r^{\prime 2} dr^\prime\\
         &=  4 \pi \rho_0 R_s^3 \left[\ln(1+r/R_s) - (1 + R_s/r)^{-1}\right].
\end{split}
\end{equation}
Here $\rho_0$ and $R_s$ are respectively the normalization and scale radius of the NFW profile.
The use of an NFW profile is empirically motivated by simulations of dark matter halos, 
and we note that other mass profiles could be assumed and may in fact provide better
alternatives.
More recent theoretical studies, for instance, have indicated that the presence of baryons
can significantly modify the mass distribution of dark matter \citep{gnedin2004,rudd2008}. 

The gas mass \Mgas\ and total mass \Mtot\ can be related by defining the gas 
fraction $\fgas \equiv \Mgas(r)/\Mtot(r)$, which could be a function of 
\Mtot, $r$, $z$, and cluster merger history.
While recently shown to be poor approximation \citep[e.g.][]{vikhlinin2006,pratt2010}, 
we make the simplifying assumption that $\fgas(r)$ is a constant.
Detailed measurements of $\fgas(r)$ typically require high significance X-ray data,
which are often insufficient or entirely lacking for clusters discovered via
the SZE.  The assumption of constant $\fgas(r)$ implies that
\begin{eqnarray}
\label{eq:gasdens}
\rhogas(r) = \fgas \,\rhotot(r).
\end{eqnarray}
We assess the impact of this assumption in Section \ref{conc}.

Using Equations \ref{eq:nfw}, \ref{eq:Mtot}, \& \ref{eq:gasdens} to solve for the gravitational 
potential energy, we find
\begin{equation}
\label{eq:UgravNFW}
\begin{split}
U_g(r)= &(4 \pi \rho_0 R_s^2)^2 G \fgas\\
&\left[\frac{R_s}{2(1 + R_s/r)^{2}}  
- \int_0^{\, r}\!\frac{\ln(1+r^\prime/R_s)}{(1+r^\prime/R_s)^{2}}dr^\prime \right],
\end{split}
\end{equation}
where we have used the fact that the differential element of gravitational 
energy for a spherical shell of gas with density $\rhogas(r)$ is 
$dU_g(r) =  - G \Mtot dm/r$, where the mass of the gas shell is 
$dm = 4 \pi \rhogas(r) r^2 dr$.

Combining Equations \ref{eq:Ysph3} \& \ref{eq:UgravNFW} through the virial relation 
(Equation \ref{eq:vir}), we have
\begin{equation}
\begin{split}
&\frac{(1+1/\mu_{\rm e})}{16 \pi^2 G \fgas} \left[3\frac{m_{\rm e} c^2}{\sigT} \Ysph(r) -4 \pi r^3 \Pe(r)\right] \\
= (\rho_0 R_s^2)^2 &\left[-\frac{R_s}{2(1 + R_s/r)^{2}} 
+ \int_0^{\, r} \!\frac{\ln(1+r^\prime/R_s)}{(1+r^\prime/R_s)^{2}}dr^\prime \right].
\label{eq:YsphNFW}
\end{split}
\end{equation}
Here the $-4 \pi r^3 \Pe(r)$ term in the brackets on the left hand side of the equation is
due to the surface pressure correction in Equation \ref{eq:vir}.

Through the above relation one can find the best fit NFW profile parameters $\rho_0$ and $R_s$
for any observationally constrained $\Ysph(r)$. We apply this method to interferometric 
SZE data in Section \ref{szeobs}.

\begin{deluxetable*}{l|ccc|ccc}
\renewcommand{\arraystretch}{0.6}
\tablefontsize{\normal}
\tablecolumns{8}
\setlength{\tabcolsep}{1.6mm}
\tablecaption{$Y_{\rm sph}$ and $M_{\rm tot}$ for Each Model Tested, 
Computed within $r_{2500}$ and $r_{500}$ Assuming Constant $\fgas=0.13$}
\tablehead{
{Cluster Name}   
& {$r_{2500}$} & {$Y_{\rm sph}(r_{2500})$} & {$M_{\rm tot}(r_{2500})$}  
& {$r_{500}$}  & {$Y_{\rm sph}(r_{500})$}  & {$M_{\rm tot}(r_{500})$}\\
~Model Fit 
& (Mpc) & {($10^{-5} {\rm Mpc}^2$)} &{($10^{14} {M_\odot}$)} 
& (Mpc) & {($10^{-5} {\rm Mpc}^2$)} &{($10^{14} {M_\odot}$)} 
} 
\startdata
{A1835}& & & & & & \\[.25pc]
~N07 (this work)	& 0.63$^{+0.01}_{-0.01}$ & ~7.64$^{+0.52}_{-0.50}$ & ~4.58$^{+0.18}_{-0.19}$ 
			& 1.45$^{+0.04}_{-0.04}$ & 17.66$^{+2.87}_{-2.36}$ & 11.24$^{+1.03}_{-0.94}$ \\[.25pc]
~A10 (this work)	& 0.63$^{+0.01}_{-0.01}$ & ~7.64$^{+0.48}_{-0.50}$ & ~4.64$^{+0.18}_{-0.19}$ 
			& 1.44$^{+0.04}_{-0.04}$ & 16.21$^{+2.47}_{-2.07}$ & 10.87$^{+0.99}_{-0.91}$ \\[.25pc]
~N07+SVM (M09)     	& 0.68$^{+0.02}_{-0.02}$ & ~8.25$^{+0.81}_{-0.78}$ & ~5.64$^{+0.58}_{-0.54}$ 
			& 1.44$^{+0.11}_{-0.10}$ & 17.55$^{+3.00}_{-2.70}$ & 11.00$^{+2.68}_{-2.22}$ \\[.25pc]
~Maughan (M09)     	& 0.66$^{+0.02}_{-0.03}$ & ~7.88$^{+0.49}_{-0.72}$ & ~5.30$^{+0.53}_{-0.72}$ 
			& 1.42$^{+0.07}_{-0.05}$ & 17.41$^{+1.61}_{-0.99}$ & 10.68$^{+1.54}_{-1.01}$ \\[.25pc]
 & & & & & &  \\
{CL~J1226+3332.9}& & & & & & \\[.25pc]
~N07 (this work)	& 0.39$^{+0.01}_{-0.01}$ & ~3.34$^{+0.28}_{-0.28}$ & ~2.35$^{+0.15}_{-0.16}$ 
			& 0.94$^{+0.02}_{-0.02}$ & ~9.43$^{+0.89}_{-0.85}$ & ~6.49$^{+0.34}_{-0.34}$ \\[.25pc]
~A10 (this work)	& 0.40$^{+0.01}_{-0.01}$ & ~3.54$^{+0.28}_{-0.29}$ & ~2.53$^{+0.14}_{-0.15}$ 
			& 0.94$^{+0.02}_{-0.02}$ & ~9.17$^{+0.88}_{-0.83}$ & ~6.42$^{+0.36}_{-0.36}$ \\[.25pc]
~N07+SVM (M09)     	& 0.41$^{+0.01}_{-0.01}$ & ~3.56$^{+0.36}_{-0.36}$ & ~2.67$^{+0.29}_{-0.27}$ 
			& 0.98$^{+0.10}_{-0.07}$ & ~9.71$^{+1.58}_{-1.29}$ & ~7.37$^{+2.50}_{-1.57}$ \\[.25pc]
~\citet{maughan2007b} 	& 0.45$^{+0.01}_{-0.01}$ & ~5.04$^{+0.31}_{-0.28}$ & ~3.41$^{+0.30}_{-0.26}$ 
			& 0.89$^{+0.02}_{-0.02}$ & 10.59$^{+0.69}_{-0.68}$ & ~5.49$^{+0.46}_{-0.47}$ \\[.25pc]
 & & & & & & \\
{A1914}& & & & & & \\[.25pc]
~N07 (this work)	& 0.60$^{+0.02}_{-0.02}$ & ~4.59$^{+0.75}_{-0.65}$ & ~3.59$^{+0.30}_{-0.30}$ 
			& 1.27$^{+0.07}_{-0.07}$ & ~7.77$^{+2.29}_{-1.70}$ & ~6.88$^{+1.26}_{-1.07}$ \\[.25pc]
~A10 (this work)	& 0.59$^{+0.02}_{-0.02}$ & ~4.37$^{+0.69}_{-0.57}$ & ~3.49$^{+0.31}_{-0.29}$ 
			& 1.23$^{+0.07}_{-0.06}$ & ~6.67$^{+1.74}_{-1.25}$ & ~6.26$^{+1.08}_{-0.87}$ \\[.25pc]
~N07+SVM (M09)     	& 0.67$^{+0.04}_{-0.03}$ & ~6.29$^{+1.03}_{-0.82}$ & ~4.97$^{+0.89}_{-0.72}$ 
			& 1.25$^{+0.11}_{-0.10}$ & 11.05$^{+2.44}_{-1.91}$ & ~6.62$^{+1.90}_{-1.42}$ \\[.25pc]
~\citet{maughan2008}  	& 0.63$^{+0.02}_{-0.02}$ & ~5.69$^{+0.37}_{-0.38}$ & ~4.31$^{+0.43}_{-0.33}$ 
		 	& 1.29$^{+0.07}_{-0.06}$ & 10.78$^{+1.03}_{-1.09}$ & ~7.49$^{+1.29}_{-1.00}$ \\[-.5pc]
& & & & & &
\enddata
\label{table:derivedQuants}
\end{deluxetable*}

\section{Application to Observations with the Sunyaev--Zel'dovich Array}\label{szeobs}

We test here the application of Equation \ref{eq:YsphNFW} to the SZE observations of the 
three clusters presented in \citet[][hereafter M09]{mroczkowski2009}, and compare
our results with the independent mass determinations presented in M09.
These three clusters span a wide range in redshift and dynamical state.
A1835, at $z=0.25$, is a relaxed, cool-core cluster \citep[e.g.,][]{peterson2001}. 
A1914, at $z=0.17$, shows evidence of being disturbed, with a
hot subcluster near the cluster core \citep{maughan2008}.
CL~J1226.9+3332 ($z=0.89$) appears somewhat relaxed given its high redshift
\citep{maughan2004b,maughan2007b}, but recent high-resolution SZE observations
with MUSTANG have indicated otherwise \citep{korngut2011}.

M09 derived mass estimates for these clusters using three data fitting methods.
The first method relied on SZE observations and X-ray surface brightness data, 
but ignored the X-ray spectroscopic data.
Instead, a density model was fit to the X-ray surface brightness data simultaneously with
a pressure profile fit to the SZE data.  Temperature information used in fitting the X-ray
surface brightness data were derived from these density and pressure profiles, assuming
the ideal gas law.  
The density model used in this method was a core-cut simplification
of that used in \citet[][hereafter V06]{vikhlinin2006}, to which
we refer as the ``Simplified V06 Model'' (SVM).  
The pressure profile used in this method is an analytic parameterization of 
the cluster radial pressure profile proposed by \citet[][hereafter N07]{nagai2007b}, 
\begin{equation}
\Pe(r) = \frac{P_{e,i}}{(r/r_p)^c 
\left[1+(r/r_p)^a\right]^{(b-c)/a}}.
\label{eq:press}
\end{equation}
This profile has the form of a generalized NFW profile, and was fit with
the slopes fixed at the best fit values found in N07, which are $(a,b,c)=(0.9,5.0,0.4)$.\footnote{
The parameters published in N07 are $(a,b,c)=(1.3,4.3,0.7)$, but the combination $(a,b,c)=(0.9,5.0,0.4)$
was later found to provide a better fit.  The planned erratum to N07 is yet to be published, 
and the corrected parameterization first appeared in M09.
However, the corrected parameterization we use here has come to be known as the 
``Nagai 07 Profile Pressure,'' and we adhere to this terminology.}
We refer to this X-ray+SZE method, which does not use or require
X-ray spectroscopic data, as the ``N07+SVM'' method.

The second mass estimation method presented in M09 was an independent, X-ray only analysis 
performed by Ben Maughan following the methods outlined in 
\citet{maughan2007b,maughan2008}.  This state of the art  
method relies on deep {\it Chandra} X-ray observations, fitting both the spectroscopic and 
surface brightness data with the full density and temperature parameterizations in V06. 
The X-ray only analyses of each cluster in M09 were published in three separate 
papers: CL~J1226.9+3332 was published in \citet{maughan2007b}, A1914 was
published in \citet{maughan2008}, and A1835 was published in M09.

The third method fit the SZE and X-ray data jointly, but relied on the assumption of 
isothermality and was included in M09 only for comparison with earlier works.
We ignore this method here, noting that its results were consistent at small radii
but increasingly discrepant at large radii.

With the density profile and the temperature or pressure profile from the above methods
in hand, the equation of hydrostatic equilibrium (Equation \ref{eq:hse}) was used to solve 
for each cluster's total mass.
The results of the N07+SVM and X-ray only analyses are reproduced in Table~\ref{table:derivedQuants}.

We compare the results of the M09 mass analyses with those from the SZE-only method presented here.
We assume the same \LCDM\ cosmology used in M09 ($\Omega_M=0.3$, $\Omega_\Lambda=0.7$, and $h=0.7$).
As in M09, we adopt the N07 profile (Equation \ref{eq:press} with $(a,b,c)=(0.9,5.0,0.4)$)
and that from \citet[][hereafter A10, Equation \ref{eq:press} with 
$(a,b,c)=(1.0510,5.4905,0.3081)$]{Arnaud2010}.

The SZE data used here were taken with the Sunyaev--Zel'dovich Array (SZA). 
Briefly, the SZA is an eight element compact array built to image the SZE in clusters 
through observations at 30 and 90~GHz \citep[see, e.g.,][]{muchovej2007}.
At 30~GHz, the SZA is sensitive to radial scales 1$'$--6$'$ over a 10$'\!$.6 diameter field of view.
At 90~GHz, the SZA measures 20$''$--120$''$ radial scales over a 3$'\!$.5 diameter field of view.
These observations are naturally fit in $u,v$-space (Fourier space) using a Markov Chain Monte
Carlo (MCMC) process, as discussed in M09.  The trial model is computed in the plane of the sky
by integrating Equation \ref{eq:compy}, using Equation \ref{eq:press} to describe the pressure.
We present here the results assuming both the N07 and A10 parameterizations.  
The sky-plane model is then Fourier transformed for direct comparison 
with the interferometric data.  This has the advantage that the likelihood of the model fit 
is computed in a basis where error bars are Gaussian.
The full details of this method can be found in M09.

The resulting fit N07 and A10 pressure profiles are used to solve for the radial profile of 
$\Ysph(r)$ (Equation \ref{eq:Ysph}).
For each accepted link in the MCMC, we fit the function described by Equation \ref{eq:YsphNFW}
assuming a constant $\fgas=0.13$ and $\mu_{\rm e}=1.17$. 
The resulting mass profiles, computed using Equation \ref{eq:Mtot}, are used to find $r_{\Delta}$
and $\Mtot(r_{\Delta})$, which are respectively the radius within which the average
density is $\Delta$ times greater than the critical density of the universe at that
redshift, and the total mass contained within that radius.  
As in M09, we report $r_{\Delta}$, $\Mtot(r_{\Delta})$, and $\Ysph(r_{\Delta})$ for
$\Delta=[2500,500]$, with statistical error bars, in Table~\ref{table:derivedQuants}.
We discuss systematics along with our conclusions in Section \ref{conc}.

\section{Conclusions}\label{conc}

We conclude that this method is remarkably consistent --- given the simplifying assumptions required 
to derive total mass using the virial relation and SZE data alone --- with the X-ray only
and X-ray+SZE mass determination methods in M09 (see Table~\ref{table:derivedQuants}). 
The assumption of constant $\fgas=0.13$ has perhaps the largest systematic impact
on the derived values of $\Mtot(r_{\Delta})$ and $r_{\Delta}$ for overdensity $\Delta$.  
The radial mass profile $\Mtot(r)$ is $\propto \fgas^{-1/2}$, as can be seen by examining 
the relation between the NFW parameter $\rho_0$ and \fgas\ in Equation \ref{eq:UgravNFW}.
However, any change in $\Mtot(r)$ affects $r_{\Delta}$ and therefore $\Mtot(r_{\Delta})$,
so the systematic change in the mass at fixed overdensity is larger than a simple
rescaling by the inverse square root of the ratio of the correct to the assumed \fgas.
Fitting the same data with an assumed $\fgas=0.11$, for example, increases $\Mtot(r_{\Delta})$
by an average of 12\% (rather than the 9\% change in the profile $\Mtot(r)$).

The assumption that $\mu_{\rm e}=1.17$, by contrast, can be expected to have a much 
smaller impact on the mass determination method presented here.
For typical abundance gradients due to metal enrichment, $\mu_{\rm e}$ varies on the 
$\sim 1\%$ level, which changes the $(1 + 1/\mu_{\rm e})$ factor in Equation \ref{eq:YsphNFW} 
on the $\sim 0.5\%$ level.  
Large systematic deviations in metallicity therefore affect the fit $\Mtot(r)$ at the $\sim 0.25\%$ level. 
The assumption of a single, constant metallicity is also common in X-ray studies of high
redshift clusters, where the limited number of X-ray counts is insufficient to constrain
more than a single spectroscopic bin.
On the other hand, helium sedimentation in the absence of magnetic fields, in a cluster
undisturbed for 3~Gyrs, could increase $\mu_{\rm e}$ in the core region by $\sim 5\%$ \citep{peng2009}.
Using the results of \citet{peng2009}, we note that the sedimentation of helium 
has little effect on the average $\mu_{\rm e}$ or on $\mu_{\rm e}(r)$ at large 
radii.\footnote{In fact, the effect of helium sedimentation is greater for X-ray surface 
brightness data than for SZE data. The redistribution of helium nuclei into the core greatly 
increases the X-ray emissivity, impacting both the charge and number of ions, and the number 
of electrons, the product of which determines the bremsstrahlung-dominated X-ray emission. 
The intensity of the SZE is only impacted by a factor proportional to the increase in the 
number of electrons. Therefore, the mass determinations from X-ray data can be expected 
to be more biased by the effects of helium sedimentation than those based on the SZE.
For good reviews, see \citet{markevitch2007,peng2009}.} 
The expectation is that mergers and magnetic fields will both suppress helium 
sedimentation \citep{peng2009}.

Another potential source of bias is due to uncertainties in the calibration 
of the SZE data.  As discussed in \citet{muchovej2007}, the absolute calibration of 
SZA data is known to better than 10\%, and the variation from observation to observation 
in amplitude of a flux calibrator (in this case Mars) is $\lesssim 5 \%$.
Calibration errors would result in scalar systematic errors in the fit pressure profile
and have a linear impact on \Ysph\ (Equation \ref{eq:Ysph}).
Examining the relation between \Ysph\ and the NFW parameter $\rho_0$ in Equation \ref{eq:YsphNFW},
we can see errors in the derived $\Mtot(r) \propto \Ysph(r)^{1/2}$, and will impact the
mass at fixed overdensity $\Mtot(r_{\Delta})$ on the $\lesssim 5\%$ level.

As noted in M09, SZA 30~GHz observations are sensitive to radial angular scales 
$\sim$1--6$'$, so the largest scale measured is $\approx r_{500}$ for A1835, and 
is $\sim 0.8 r_{500}$ for A1914. The values we report should therefore be treated as 
extrapolations of the fit, and depend on the assumed N07 or A10 pressure profile.

{The surface pressure term in the virial relation (Equation \ref{eq:vir}) was not included in the
original work, and can be of the same order as the thermal energy of the cluster gas.
While this term systematically reduces the mass inferred through this method, the amount
the estimate changed from the value neglecting surface pressure was typically within 
the error bars.  Overall, the inclusion of this term was found to improve the agreement
between the SZE-only inferred mass and the estimates from X-ray and SZE+X-ray.}

Other systematics the SZE-only mass estimation method could suffer are common to X-ray mass
determinations that rely on a cluster's fit radial temperature and density 
profiles and assume thermal HSE to derive mass.
First, HSE is most readily applied by assuming spherical symmetry.
Second, while the gas may be virialized within the potential and supported 
predominantly by thermal pressure, there is an expectation that $\sim$10\%--20\% of the total
pressure is due to turbulent motions in the ICM \citep{lau2009, battaglia2010}.  
This is equivalent to including kinetic energy, in addition to thermal, in the virial relation.
The interesting question arises as to whether adopting the virial relation, instead of 
just HSE, could be used to make X-ray mass determinations more robust.

The broader implications of this work are that by solving for the radial thermal energy profile
one can estimate cluster mass from any SZE observation that can constrain 
that cluster's radial pressure profile.  Radial profiles have already been fit to 
clusters observed by ACT, SPT, and now Planck \citep{menanteau2010,plagge2010,Planck2011scaling}, 
and this method may prove particularly useful for providing initial mass estimates 
for the clusters they discover.  
A future work, using SZA observations of a more complete sample of clusters,
will compare total mass derived in this way with X-ray and lensing mass estimates.
We will also consider how to extend this method for the case where the gas fraction
varies with cluster radius.

\begin{acknowledgements}
The author wishes to acknowledge the diligence of, and help from, Colin Hill, who
brought to his attention the neglect of the surface pressure term.  
The author thanks Erik Reese for many useful discussions 
and Marshall Joy for encouraging him to test this method on a handful of clusters.
The author also thanks the anonymous referee for comments that helped
improve the direction and focus of this Letter.
Support for the author was provided by NASA through the Einstein Fellowship Program, 
grant PF0-110077.
\end{acknowledgements}

{\it Facilities:} \facility{SZA}



\end{document}